\documentclass[osajnl,twocolumn,showpacs,floatfix,bibnotes]{revtex4}

\usepackage{here}

\usepackage{graphicx}

\newcommand\pictc[5]{\begin{figure}
                       \centerline{
                       \includegraphics*[width=#1\columnwidth,]{#3}}
                   \protect\caption{\protect\label{fig:#4} #5}
                    \end{figure}            }
\newcommand\pict[4][0.95]{\pictc{#1}{!tb}{#2}{#3}{#4}}
\newcommand\rpict[1]{\ref{fig:#1}}

\newcommand\leqt[1]{\protect\label{eq:#1}}
\newcommand\reqtn[1]{\ref{eq:#1}}
\newcommand\reqt[1]{(\reqtn{#1})}

\newcounter{Fig}

\begin{document}
\begin{sloppy}

\title{Soliton switching and Bloch-wave filtering in periodic photonic lattices}

\author{Andrey A. Sukhorukov}
\author{Yuri S. Kivshar}
\affiliation{Nonlinear Physics Centre and Centre for Ultrahigh
bandwidth Devices for Optical Systems (CUDOS), Research School of
Physical Sciences and Engineering, Australian National University,
Canberra, ACT 0200, Australia}
\homepage{http://www.rsphysse.anu.edu.au/nonlinear}

\begin{abstract}
We suggest a novel concept in the soliton switching based on the
Bloch-wave filtering in periodic photonic structures. Taking a
binary waveguide array as an example, we demonstrate that spatial
solitons that belong to different spectral band gaps can
be selectively reflected from or transmitted through an engineered
defect, which acts as a low or high-pass filter for Bloch waves.
\end{abstract}


\maketitle

The study of nonlinear light propagation in periodic photonic
structures recently attracted strong interest because it presents
the unique possibility of observing experimentally an interplay
between the effects of nonlinearity and
periodicity~\cite{Christodoulides:2003-817:NAT}. Fabricated
nonlinear periodic structures such as arrays of coupled nonlinear
optical waveguides can simultaneously support distinctive types of
self-trapped optical beams in the form of discrete and gap spatial
optical solitons~\cite{Kivshar:2003:OpticalSolitons}.

{\em Discrete solitons} in a self-focusing nonlinear medium are
localized due to the effect of total internal reflection, which
allows to trap an optical beam in a few neighboring waveguides
suppressing diffraction-induced light
spreading~\cite{Christodoulides:1988-794:OL,
Eisenberg:1998-3383:PRL}. Spatial {\em gap solitons} can also
exist in the same periodic structures, and they appear due to the
Bragg scattering experienced by waves incident on a periodic
lattice at a particular angle~\cite{Mandelik:2004-93904:PRL,
Neshev:2004-83905:PRL}.

Many schemes of all-optical soliton switching suggested so far are
based on the use of the specific properties of discrete solitons~\cite{Morandotti:2003-834:OL, Christodoulides:2003-817:NAT}.
Here, we suggest a novel approach which allows one to switch
selectively solitons from different gaps. Our idea is related to a
more general concept of {\em Bloch-wave filtering}, and it is
based on the observation that a structural defect introduced into
a periodic photonic lattice can be designed in such a way that it
would allow transmission of Bloch waves from some bands
simultaneously reflecting all waves that belong to other bands. As
a result, the gap solitons associated with those spectral bands
may also be transmitted or reflected selectively. In this Letter,
we demonstrate that the concept of the Bloch-wave filtering can be
realized very effectively in a binary array of weakly coupled
optical waveguides recently demonstrated and studied
experimentally~\cite{Morandotti:2004-2890:OL}.

We consider spatial solitons in nonlinear periodic structures such
as waveguide arrays and one-dimensional photonic lattices. In
general, propagation of light in such systems can be described by
a parabolic wave equation for the normalized electric field
envelope $E(x,z)$,
\begin{equation} \leqt{NLS}
   i \frac{\partial E}{\partial z}
   + D \frac{\partial^2 E}{\partial x^2}
   + \nu(x) E + {\cal F}(x, |E|^2) E = 0,
\end{equation}
where $x$ and $z$ are the transverse and propagation coordinates
normalized to the characteristic values $x_s$ and $z_s$,
respectively, $D = z_s \lambda / (4 \pi n_0 x_s^2)$ is the beam
diffraction coefficient, $n_0$ is the average medium refractive
index, $\lambda$ is the vacuum wavelength, $\nu(x) = 2 \Delta n(x)
\pi n_0 / \lambda$, $\Delta n(x) = \Delta n(x+d)$ is the effective
modulation of the optical refractive index with the spatial period
$d$, and ${\cal F}(x, |E|^2) = {\cal F}(x+d, |E|^2)$ is the periodic nonlinear response function.

Linear waves propagating through a periodic lattice are the 
Bloch modes $E_{\kappa,n}(x,z) = \psi_{\kappa,n}(x) \exp( i \kappa
x / d + i \beta_{\kappa,n} z )$ satisfying periodicity condition
$\psi_{\kappa,n}(x) = \psi_{\kappa,n}(x+d)$. Real wave-numbers
$\kappa$ describe propagating Bloch waves, which dispersion curves
$\beta_{\kappa,n}(\kappa)$ form distinct bands with the indices
$n=1,2,\ldots$. Nonlinear dynamics of slowly modulated Bloch
waves of the form $E(x,z) = A(x,z) E_{\kappa,n}(x,z)$, can be described by an effective nonlinear Sch\"odinger equation~\cite{Sipe:1988-132:OL}
for the complex envelope function $A(x,z)$,
\[
  i \frac{\partial A}{\partial z}
  + i V_{\kappa,n} \frac{\partial A}{\partial x} 
  + D_{\kappa,n} \frac{\partial^2 A}{\partial x^2} 
  + \gamma_{\kappa,n} |A|^2 A 
  = 0.
\]
Here, $V_{\kappa,n} = -d \beta_{\kappa,n} / d \kappa$ defines the group velocity or the beam propagation angle, $D_{\kappa,n} = -d^2 \beta_{\kappa,n} / d \kappa^2$ is coefficient of diffraction experienced by the beam, and $\gamma_{\kappa,n} = \int_0^d {\cal F}(x, |E_{\kappa,n}|^2)
|E_{\kappa,n}|^2 \, dx / \int_0^d |E_{\kappa,n}|^2 \, dx$ is the
effective nonlinear coefficient.

\pict{fig01.eps}{solitons}{ (a)~Refractive index profile in a
binary waveguide array; (b)~Bloch wavenumber vs. the propagation
constant (bands are shaded); (c,d)~profiles of gap and discrete
solitons associated with bands 2 and 1, respectively. The
parameters are $d_n=2.5\mu$m, $d_w=4\mu$m, $d_s=5\mu$m. }

Nonlinear self-action can suppress diffraction-induced beam
spreading leading to the formation of a lattice soliton when
$D_{\kappa,n} \gamma_{\kappa,n} > 0$. Such soliton has the form $A = A_0 {\rm sech}[(x-V_{\kappa,n} z)/x_0] \exp(i \rho z)$, where $A_0=\sqrt{2\rho/\gamma_{\kappa,n}}$ and
$x_0=\sqrt{D_{\kappa,n}/\rho}$. Most remarkably, solitons may occur in both self-focusing
($\gamma>0$) and self-defocusing ($\gamma<0$) nonlinear media
since Bloch waves can exhibit either normal ($D_{\kappa,n}>0$) or
anomalous ($D_{\kappa,n}<0$) diffraction in each of the bands.
Therefore, a nonlinear periodic structure can support
simultaneously different types of solitons associated with various
bands~\cite{Pelinovsky:2004-36618:PRE}.

To be specific, we consider the case of self-focusing Kerr-type
nonlinearity [${\cal F}(x,I) = I$] in binary waveguide arrays, corresponding to the AlGaAS superlattices where the formation of two types of optical solitons has recently been observed
experimentally~\cite{Morandotti:2004-2890:OL}. The 
refractive index profile of the optical superlattice is shown in
Fig.~\rpict{solitons}(a), and other paremeters are $n_0=3.3947$ and $\lambda=1.5\,\mu m$. Solitons appear in the spectral regions
with normal diffraction, and at larger intensities their
propagation constants are increased and moved into the gaps, as
illustrated by arrows in Fig.~\rpict{solitons}(b). Solitons
associated with the first and second bands are mainly confined at
wide and narrow waveguides, respectively [see
Figs.~\rpict{solitons}(c,d)], reflecting the structure of
corresponding Bloch waves.

\pict{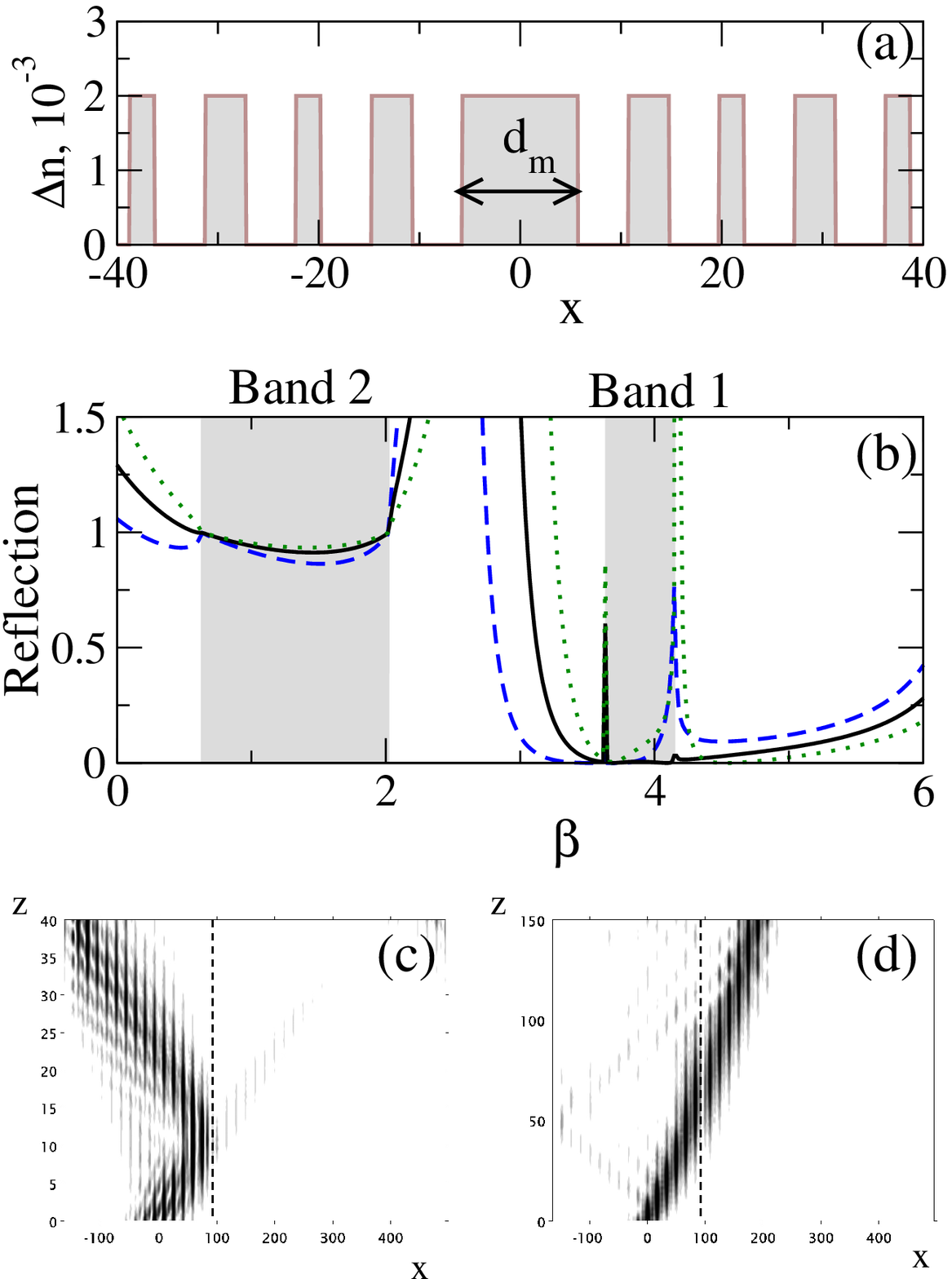}{reflectN}{ (a)~Refractive index profile of
waveguide array with a defect created by increasing the size of a
single narrow waveguide; (b)~Bloch-wave reflection coefficients
for different defect widths: $d_m=11\mu m$ (dashed), $11.5 \mu m$
(solid), and $12 \mu m$ (dotted); (c,d)~Numerical simulations of
(c)~band-2 and (d)~band-1 soliton interactions with engineered
defect corresponding to the solid curve in~(b). }

\pict{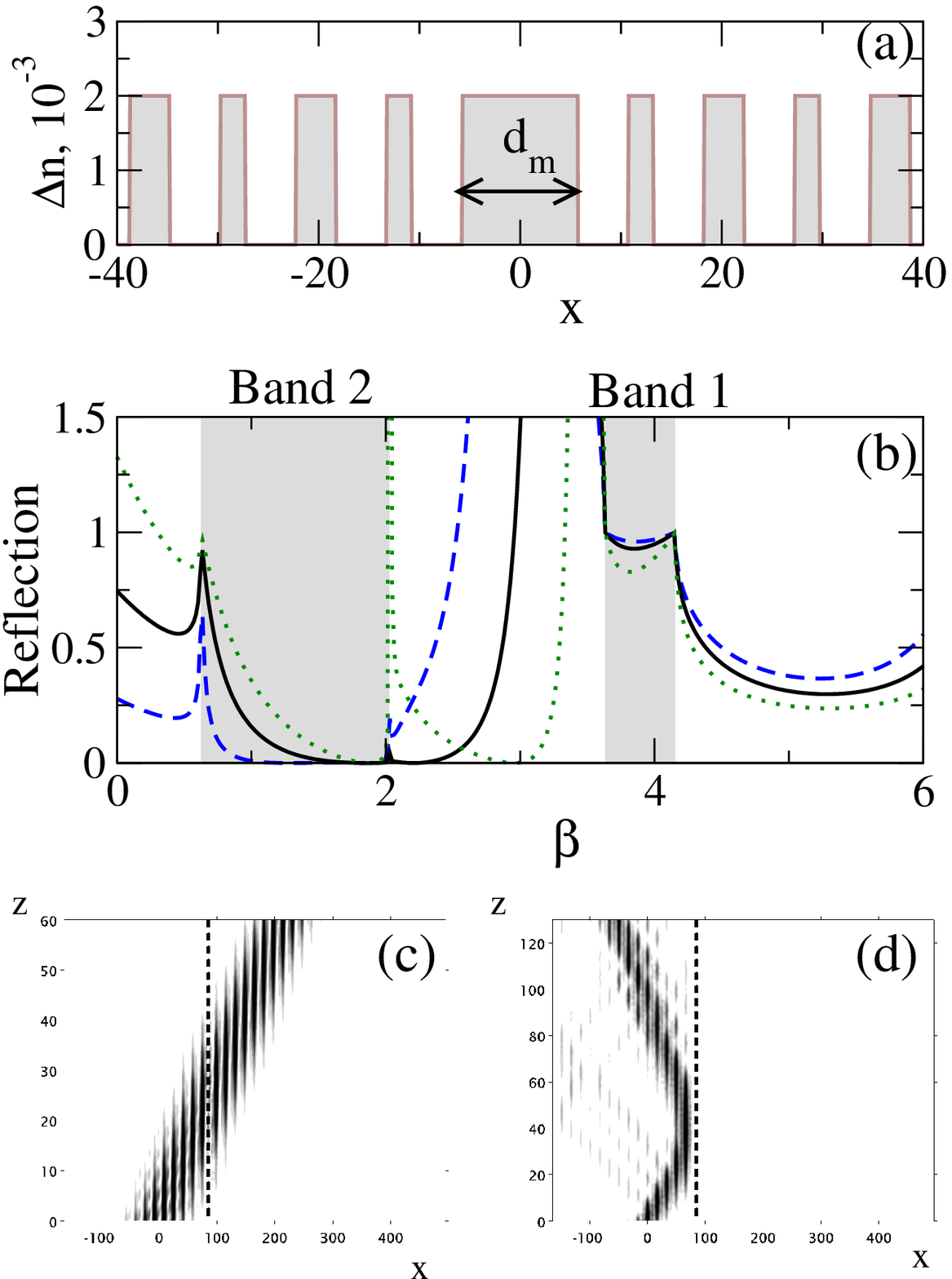}{reflectW}{ Results for a a defect created by
increasing the size of a single wide waveguide. All notations and
parameters correspond to Fig.~\rpict{reflectN}. }

We now study the wave transmission through a defect created in a
waveguide array. The soliton-defect scattering in a homogeneous
waveguide array has been studied experimentally in
Ref.~\cite{Morandotti:2003-834:OL}, but here we are interested in
multi-gap effects and take the case of a binary waveguide array as
the simplest example of such a system.

If a wave with a particular wave-number $\kappa(\beta)$ is
incident on a defect from the left, then the field structure can
be represented as a superposition of Bloch waves,
\begin{equation} \leqt{defect}
  \begin{array}{l} {\displaystyle
     E = a_i E_{\kappa,n}(x,z) + a_r E_{-\kappa,n}(x,z), \;\; x < x_{d-},
   } \\*[9pt]  {\displaystyle
     E = a_t E_{\kappa,n}(x,z), \;\; x > x_{d+},
   } \end{array}
\end{equation}
where $a_i$, $a_r$, and $a_t$ are the amplitudes of the incident,
reflected, and transmitted waves, respectively, and $x_{d-} < x <
x_{d+}$ is the defect extension. The key idea of our approach is
to design a defect which reflects strongly the waves belonging to
one band, but transmits almost completely the waves belonging to
another band. For the binary superlattice, we find that such
Bloch-wave filtering can be realized by {\em increasing} the width
of a single narrow or wide waveguide, as shown in
Fig.~\rpict{reflectN}(a) and Fig.~\rpict{reflectW}(a). We plot the
corresponding dependencies of the reflection coefficients
$|a_r/a_i|^2$ on the propagation constant $\beta$ for different
defect widths in Fig.~\rpict{reflectN}(b) and
Fig.~\rpict{reflectW}(b). When the width of originally narrow
[Fig.~\rpict{reflectN}] or wide [Fig.~\rpict{reflectW}]
waveguide is increased to a particular value, the propagation constants of the second-order mode of the defect waveguide and fundamental modes of neighboring waveguides almost coincide, allowing for an efficient field tunneling between the waveguides. Simultaneously, the tunneling is suppressed between the next-neighbor waveguides which have different propagation constants. Since Bloch waves are mainly confined at wide and narrow waveguides in the first and second bands, their resonant transmission can be supported by a defect embedded between the wide [Fig.~\rpict{reflectN}] or narrow [Fig.~\rpict{reflectW}] waveguides, respectively, whereas waves are fully reflected in the complimentary gaps.

As discussed above, in a self-focusing medium the soliton wavenumbers
are located in the upper sections of the bands, but can also be shifted
inside the gap. Therefore, to realize efficient soliton
transmission through a defect, it is necessary to reduce
reflection not only inside a band, but also in the gap outside the
band. Bloch wave profiles in the gap represent the tails of
solitons and decay inside the structure. For such evanescent waves
the reflection coefficients may be larger than unity, and even attain infinite values. The latter case corresponds to the existence of localized modes supported by the defect, i.e. solutions with $a_i = 0$ and $a_r,a_t \ne 0$. Resonant coupling between the soliton and linear defect modes can result in power transfer to the localized state. Optimal soliton transmission can be achieved when such resonances are avoided, i.e. when the reflection coefficient in the vicinity of the upper band edge is close to zero. This is realized for defect widths corresponding to solid curves in Fig.~\rpict{reflectN}(b) and Fig.~\rpict{reflectW}(b).
Additionally, we have verified that the phase of the transmitted waves
is almost constant throughout the band, minimizing distortions to
the transmitted soliton profiles due to spatial group-velocity
dispersion.

Our numerical simulations of Eq.~\reqt{NLS} confirm that the designed Bloch-wave
filters can switch solitons very efficiently. As initial conditions, we have used solutions for stationary solitons, and induced their motion across the lattice by inclining the phase front in the transverse direction. We note that controlled generation of band-1 and band-2 solitons is readily accessible experimentally (see Ref.~\cite{Morandotti:2004-2890:OL} and references therein).
We have verified that the solitons are always reflected from a defect
which inhibits transmission of the corresponding Bloch waves, see
examples in Fig.~\rpict{reflectN}(c) and Fig.~\rpict{reflectW}(d).
On the other hand, solitons from the complimentary bands are
transmitted through the defect with a minimal amount of radiation
if their velocities exceed critical values, see
Fig.~\rpict{reflectN}(d) and Fig.~\rpict{reflectW}(c). Strongly
localized solitons traveling at slow velocities may however be
reflected back, due to substantial modification of defect response through nonlinear self-action. This behavior is consistent with general features of soliton-defect interactions.

In conclusion, we have suggested a novel approach to the soliton
switching in periodic photonic structures and demonstrated a
simple design of highly efficient Bloch filters for the example of a
binary optical superlattice. The concept of Bloch-wave filtering
is rather general, and it can be applied to various types of
periodic structures such as  optically-induced nonlinear lattices
and photonic crystals. Application of this concept to
two-dimensional discrete networks will allow to greatly enhance
their potential for all-optical signal manipulation.

The authors acknowledge a support from the Australian Research
Council and thank B. Eggleton and M. de Sterke for useful
discussions.

\end{sloppy}
\end{document}